\documentclass[aps,manuscript,preprint,onecolumn,aps,pre,eqsecnum,longbibliography,showpacs,showkeys,a4paper]{revtex4-1}
\usepackage[T1]{fontenc}
\usepackage[latin9]{inputenc}
\usepackage{color}
\definecolor{note_fontcolor}{rgb}{0.800781, 0.800781, 0.800781}
\usepackage[english]{babel}
\usepackage{units}
\usepackage{amsmath}
\usepackage{amssymb}
\usepackage[unicode=true,
 bookmarks=true,bookmarksnumbered=true,bookmarksopen=true,bookmarksopenlevel=1,
 breaklinks=true,pdfborder={0 0 0},backref=false,colorlinks=true]
 {hyperref}
\hypersetup{
 pdfauthor={AINS}}

\makeatletter

\newenvironment{lyxgreyedout}
  {\textcolor{note_fontcolor}\bgroup\ignorespaces}
  {\ignorespacesafterend\egroup}

\@ifundefined{textcolor}{}
{%
 \definecolor{BLACK}{gray}{0}
 \definecolor{WHITE}{gray}{1}
 \definecolor{RED}{rgb}{1,0,0}
 \definecolor{GREEN}{rgb}{0,1,0}
 \definecolor{BLUE}{rgb}{0,0,1}
 \definecolor{CYAN}{cmyk}{1,0,0,0}
 \definecolor{MAGENTA}{cmyk}{0,1,0,0}
 \definecolor{YELLOW}{cmyk}{0,0,1,0}
}
\numberwithin{equation}{section}
\numberwithin{figure}{section}
\numberwithin{table}{section}

\usepackage{graphicx}

\DeclareGraphicsExtensions{.png,.pdf,.eps}
\graphicspath{{pictures/}}

\makeatother

\begin{document}

\title{Implications of a deeper level explanation of the deBroglie--Bohm
version of quantum mechanics\vspace*{\bigskipamount}
}

\author{Gerhard \surname{Grössing}\textsuperscript{}}

\email[E-mail: ]{ains@chello.at}

\homepage[Visit: ]{http://www.nonlinearstudies.at/}

\author{Siegfried \surname{Fussy}\textsuperscript{}}

\email[E-mail: ]{ains@chello.at}

\homepage[Visit: ]{http://www.nonlinearstudies.at/}

\author{Johannes \surname{Mesa Pascasio}\textsuperscript{}}

\email[E-mail: ]{ains@chello.at}

\homepage[Visit: ]{http://www.nonlinearstudies.at/}

\author{Herbert \surname{Schwabl}\textsuperscript{}}

\email[E-mail: ]{ains@chello.at}

\homepage[Visit: ]{http://www.nonlinearstudies.at/}

\affiliation{\textsuperscript{}Austrian Institute for Nonlinear Studies, Akademiehof\\
 Friedrichstr.~10, 1010 Vienna, Austria\\
\vspace*{1cm}
}

\date{\today}
\begin{abstract}
Elements of a ``deeper level'' explanation of the deBroglie--Bohm
(dBB) version of quantum mechanics are presented. Our explanation
is based on an analogy of quantum wave-particle duality with bouncing
droplets in an oscillating medium, the latter being identified as
the vacuum's zero-point field. A hydrodynamic analogy of a similar
type has recently come under criticism by Richardson \emph{et~al}.~\cite{Richardson.2014analogy},
because despite striking similarities at a phenomenological level
the governing equations related to the force on the particle are evidently
different for the hydrodynamic and the quantum descriptions, respectively.
However, said differences are not relevant if a radically different
use of said analogy is being made, thereby essentially referring to
emergent processes in our model. If the latter are taken into account,
one can show that the forces on the particles are identical in both
the dBB and our model. In particular, this identity results from an
exact matching of our emergent velocity field with the Bohmian ``guiding
equation''. One thus arrives at an explanation involving a deeper,
i.e.\ subquantum, level of the dBB version of quantum mechanics.
We show in particular how the classically-local approach of the usual
hydrodynamical modeling can be overcome and how, as a consequence,
the configuration-space version of dBB theory for \emph{N} particles
can be completely substituted by a ``superclassical'' emergent dynamics
of \emph{N} particles in real 3-dimensional space.%
\begin{lyxgreyedout}
\noindent \global\long\def\VEC#1{\mathbf{#1}}
\global\long\def\d{\,\mathrm{d}}
\global\long\def\e{{\rm e}}
\global\long\def\meant#1{\left<#1\right>}
\global\long\def\meanx#1{\overline{#1}}
\global\long\def\mpbracket{\ensuremath{\genfrac{}{}{0pt}{1}{-}{\scriptstyle (\kern-1pt +\kern-1pt )}}}
\global\long\def\pmbracket{\ensuremath{\genfrac{}{}{0pt}{1}{+}{\scriptstyle (\kern-1pt -\kern-1pt )}}}
\global\long\def\p{\partial}
\end{lyxgreyedout}

\end{abstract}

\keywords{quantum mechanics, hydrodynamics, deBroglie--Bohm theory, guiding
equation, configuration space, zero-point field}

\maketitle

\section{Introduction\label{sec:intro}}

The Schrödinger equation for $N>1$ particles does not describe a
wave function in ordinary 3-dimensional space, but instead in an abstract
$3N$-dimensional space. For quantum realists, including Schrödinger
and Einstein, for example, this has always been considered as ``indigestible''.
This holds even more so for a realist, causal approach to quantum
phenomena such as the deBroglie--Bohm (dBB) version of quantum mechanics.
David Bohm himself has admitted this, calling it a ``serious problem'':
``While our theory can be extended formally in a logically consistent
way by introducing the concept of a wave in a $3N$-dimensional space,
it is evident that this procedure is not really acceptable in a physical
theory, and should at least be regarded as an artifice that one uses
provisionally until one obtains a better theory in which everything
is expressed once more in ordinary $3$-dimensional space.''~\cite{Bohm.1997causality}
(For more detailed accounts of this discussion already in the early
years of quantum mechanics, see~\cite{Norsen.2010theory} and \cite{Norsen.2014can}.)

In the present paper, we shall refer to our attempt towards such a
``better theory'' in terms of a deeper level, i.e.\ subquantum,
approach to the dBB theory, and thus to quantum theory in general.
In fact, with our model we have in a series of papers already obtained
several essential elements of nonrelativistic quantum theory~\cite{Groessing.2008vacuum,Groessing.2009origin,Groessing.2012doubleslit,Groessing.2013dice}.
They derive from the assumption that a particle of energy $E=\hbar\omega$
is actually an oscillator of angular frequency $\omega$ phase-locked
with the zero-point oscillations of the surrounding environment, the
latter containing both regular and fluctuating components and being
constrained by the boundary conditions of the experimental setup via
the buildup and maintenance of standing waves. The particle in this
approach is an off-equilibrium steady state oscillation maintained
by a constant throughput of energy provided by the (\textquotedblleft classical\textquotedblleft )
zero-point energy field. We have, for example, applied the model to
the case of interference at a double slit, thereby obtaining the exact
quantum mechanical probability density distributions on a screen behind
the double slit, the average trajectories (which because of the averaging
are shown to be identical to the Bohmian ones), and the involved probability
density currents. Our whole model is constructed in close analogy
to the bouncing/walking droplets above the surface of a vibrated liquid
in the experiments first performed by Yves Couder, Emmanuel Fort and
co-workers~\cite{Couder.2006single-particle,Couder.2012probabilities,Fort.2010path-memory},
which in many respects can serve as a classical prototype guiding
our intuition for the modeling of quantum systems.

However, there are also obvious differences between the mentioned
physics of classical bouncers/walkers on the one hand, and the hydrodynamic-like
models for quantum systems like our own model or the dBB one on the
other hand. In a recent paper, Richardson \emph{et~al}.~\cite{Richardson.2014analogy}
have probed more thoroughly into the hydrodynamic analogy of dBB-type
quantum wave-particle duality with that of the classical bouncing
droplets. Apart from the obvious difference in that Bohmian theory
is distinctly nonlocal, whereas droplet-surface interactions are rooted
in classical hydrodynamics and thus in a manifestly local theory,
Richardson \emph{et~al}.\ focus on the following observation: the
evidently different nature of the Bohmian force upon a quantum particle
as compared to the force that a surface wave exerts upon a droplet.
In fact, wherever the probability density in the dBB picture is close
to zero, the quantum force becomes singular and will very quickly
push any particle away from that area. Conversely, the hydrodynamic
force directs the droplet into the trough of the wave! So, the probability
of finding a droplet in the minima never reaches zero as it does for
a quantum particle. The authors conclude that these discrepancies
between the two models highlight ``a major difference between the
hydrodynamic force and the quantum force''.\ \cite{Richardson.2014analogy} 

Although these authors generally recover in numerical hydrodynamic
simulations the results of the Paris group (later confirmed also by
the group of John Bush at MIT~\cite{Bush.2015pilot-wave}) on single-slit
diffraction and double-slit interference, they also point out the
(already known) striking contrast between the trajectory behaviors
for the bouncing droplet systems and dBB-type quantum mechanics, respectively.
Whereas the latter exhibits the well-known no-crossing property, the
trajectories of the former do to a large extent cross each other.
So, again, the physics in the two models is apparently fundamentally
different, despite some striking similarities on a phenomenological
level. As to the differences, one may very well expect that they will
even become more severe when moving from one-particle to $N$-particle
systems.

So, all in all, the paper by Richardson \emph{et~al}.~\cite{Richardson.2014analogy}
cautions against the assumption of too close a resemblance of bouncer/walker
systems and the hydrodynamic-like modeling of quantum systems like
the dBB one, with their main argument being that the hydrodynamic
force on a droplet strikingly contrasts with the quantum force on
a particle in the dBB theory. However, we shall here argue against
the possible conclusion that one has thus reached the limits of applicability
of the hydrodynamic bouncer analogy for quantum modeling. On the contrary,
as we have already pointed out in previous papers, it is a more detailed
model inspired by the bouncer/walker experiments that can show the
fertility of said analogy. It enables us to show that our model, being
of the type of an ``emergent quantum mechanics''~\cite{Groessing.2012emerqum11-book,Groessing.2014emqm13-book},
can provide a deeper-level explanation of the dBB version of quantum
mechanics (Chapter~2). Moreover, as we shall also show, it turns
out to provide an identity of an emergent force on the bouncer in
our hydrodynamic-like model with the quantum force in Bohmian theory
(Chapter~3). Finally, in Chapter~4 we shall discuss the ``price''
to be paid in order to arrive at our explanation of dBB theory in
that some kind of nonlocality, or a certain ``systemic nonlocality'',
has to be admitted in the model from the start. However, the simplicity
and elegance of our derived formalism, combined with arguments about
the reasonableness of a corresponding hydrodynamic-like modeling,
will show that our approach may be a viable one w.r.t.\ understanding
the emergence of quantum phenomena from the interactions and contextualities
provided by the combined levels of classical boundary conditions and
those of a subquantum domain.

\section{Identity of the emergent kinematics of \emph{$N$} bouncers in real
$3$-dimensional space with the configuration-space version of deBroglie--Bohm
theory for \emph{$N$} particles \label{sec:config}}

Consider one particle in an $n$-slit system. In quantum mechanics,
as well as in our quantum-like modeling via an emergent quantum mechanics
approach, one can write down a formula for the total intensity distribution
$P$ which is very similar to the classical formula. For the general
case of $n$ slits, it holds with phase differences $\varphi_{ij}=\varphi_{i}-\varphi_{j}$
that
\begin{equation}
P=\sum_{i=1}^{n}\left(P_{i}+\sum_{j=i+1}^{n}2R_{i}R_{j}\cos\varphi_{ij}\right),\label{eq:Sup2.1}
\end{equation}
where the phase differences are defined over the whole domain of the
experimental setup. Apart from the role of the relative phase with
important implications for the discussions on nonlocality~\cite{Groessing.2013dice},
there is one additional ingredient that distinguishes~(\ref{eq:Sup2.1})
from its classical counterpart, namely the ``dispersion of the wavepacket''.
As in our model the ``particle'' is actually a bouncer in a fluctuating
wave-like environment, i.e.~analogously to the bouncers of Couder
and Fort's group, one does have some (e.g.\ Gaussian) distribution,
with its center following the Ehrenfest trajectory in the free case,
but one also has a diffusion to the right and to the left of the mean
path which is just due to that stochastic bouncing. Thus the total
velocity field of our bouncer in its fluctuating environment is given
by the sum of the forward velocity $\VEC v$ and the respective diffusive
velocities $\VEC u_{\mathrm{L}}$ and $\VEC u_{\mathrm{R}}$ to the
left and the right. As for any direction $i$ the diffusion velocity
$\VEC u_{\mathrm{i}}=D\frac{\nabla_{i}P}{P}$ does not necessarily
fall off with the distance, one has long effective tails of the distributions
which contribute to the nonlocal nature of the interference phenomena~\cite{Groessing.2013dice}.
In sum, one has three, distinct velocity (or current) channels per
slit in an $n$-slit system. 

We have previously shown~\cite{Fussy.2014multislit,Groessing.2014relational}
how one can derive the Bohmian guidance formula from our bouncer/walker
model. To recapitulate, we recall the basics of that derivation here.
Introducing classical wave amplitudes $R(\VEC w_{i})$ and generalized
velocity field vectors $\VEC w_{i}$, which stand for either a forward
velocity $\VEC v_{i}$ or a diffusive velocity $\VEC u_{i}$ in the
direction transversal to $\VEC v_{i}$, we account for the phase-dependent
amplitude contributions of the total system's wave field projected
on one channel's amplitude $R(\VEC w_{i})$ at the point $(\VEC x,t)$
in the following way: We define a \emph{conditional probability density}
$P(\VEC w_{i})$ as the local wave intensity $P(\VEC w_{i})$ in one
channel (i.e.~$\VEC w_{i}$) upon the condition that the totality
of the superposing waves is given by the ``rest'' of the $3n-1$
channels (recalling that there are 3 velocity channels per slit).
The expression for $P(\VEC w_{i})$ represents what we have termed
``relational causality'': any change in the local intensity affects
the total field, and \emph{vice versa}, any change in the total field
affects the local one. In an $n$-slit system, we thus obtain for
the conditional probability densities and the corresponding currents,
respectively, i.e.\ for each channel component $\mathit{i}$,
\begin{align}
P(\VEC w_{i}) & =R(\VEC w_{i})\VEC{\hat{w}}_{i}\cdot{\displaystyle \sum_{j=1}^{3n}}\VEC{\hat{w}}_{j}R(\VEC w_{j})\label{eq:Proj-1}\\
\VEC J\mathrm{(}\VEC w_{i}\mathrm{)} & =\VEC w_{i}P(\VEC w_{i}),\qquad i=1,\ldots,3n,
\end{align}
with
\begin{equation}
\cos\varphi_{i,j}:=\VEC{\hat{w}}_{i}\cdot\VEC{\hat{w}}_{j}.
\end{equation}
Consequently, the total intensity and current of our field read as
\begin{align}
P_{\mathrm{tot}}= & {\displaystyle \sum_{i=1}^{3n}}P(\VEC w_{i})=\left({\displaystyle \sum_{i=1}^{3n}}\VEC{\hat{w}}_{i}R(\VEC w_{i})\right)^{2}\label{eq:Ptot6-1}\\
\VEC J_{\mathrm{tot}}= & \sum_{i=1}^{3n}\VEC J(\VEC w_{i})={\displaystyle \sum_{i=1}^{3n}}\VEC w_{i}P(\VEC w_{i}),\label{eq:Jtot6-1}
\end{align}
 leading to the \textit{emergent total velocity}
\begin{equation}
\VEC v_{\mathrm{tot}}=\frac{\VEC J_{\mathrm{tot}}}{P_{\mathrm{tot}}}=\frac{{\displaystyle \sum_{i=1}^{3n}}\VEC w_{i}P(\VEC w_{i})}{{\displaystyle \sum_{i=1}^{3n}}P(\VEC w_{i})}\,.\label{eq:vtot_fin-1}
\end{equation}

In~\cite{Fussy.2014multislit,Groessing.2014relational} we have shown
with the example of $n=2,$ i.e.\ a double slit system, that Eq.~(\ref{eq:vtot_fin-1})
can equivalently be written in the form
\begin{equation}
\VEC v_{\mathrm{tot}}=\frac{R_{1}^{2}\VEC v_{\mathrm{1}}+R_{2}^{2}\VEC v_{\mathrm{2}}+R_{1}R_{2}\left(\VEC v_{\mathrm{1}}+\VEC v_{2}\right)\cos\varphi+R_{1}R_{2}\left(\VEC u_{1}-\VEC u_{2}\right)\sin\varphi}{R_{1}^{2}+R_{2}^{2}+2R_{1}R_{2}\cos\varphi}\,.\label{eq:vtot-1}
\end{equation}

The trajectories or streamlines, respectively, are obtained according
to $\VEC{\dot{x}}=\VEC v_{\mathrm{tot}}$ in the usual way by integration.
As first shown in~\cite{Groessing.2012doubleslit}, by re-inserting
the expressions for convective and diffusive velocities, respectively,
i.e.\ $\VEC v_{i}=\frac{\nabla S_{i}}{m}$, $\VEC u_{i}=-\frac{\hbar}{m}$$\frac{\nabla R_{i}}{R_{i}}$,
one immediately identifies Eq.~(\ref{eq:vtot-1}) with the Bohmian
guidance formula. Naturally, employing the Madelung transformation
for each path $j$ ($j=1$ or $2$), 
\begin{equation}
\psi_{j}=R_{j}\e^{\mathrm{i}S_{j}/\hbar},\label{eq:3.14-1}
\end{equation}
and thus $P_{j}=R_{j}^{2}=|\psi_{j}|^{2}=\psi_{j}^{*}\psi_{j}$, with
$\varphi=(S_{1}-S_{2})/\hbar$, and recalling the usual trigonometric
identities such as $\cos\varphi=\frac{1}{2}\left(\e^{\mathrm{i}\varphi}+\e^{-\mathrm{i}\varphi}\right)$,
one can rewrite the total average current immediately in the usual
quantum mechanical form as 
\begin{equation}
\begin{array}{rl}
{\displaystyle \mathbf{J}_{{\rm tot}}} & =P_{{\rm tot}}\mathbf{v}_{{\rm tot}}\\[3ex]
 & ={\displaystyle (\psi_{1}+\psi_{2})^{*}(\psi_{1}+\psi_{2})\frac{1}{2}\left[\frac{1}{m}\left(-\mathrm{i}\hbar\frac{\nabla(\psi_{1}+\psi_{2})}{(\psi_{1}+\psi_{2})}\right)+\frac{1}{m}\left(\mathrm{i}\hbar\frac{\nabla(\psi_{1}+\psi_{2})^{*}}{(\psi_{1}+\psi_{2})^{*}}\right)\right]}\\[3ex]
 & ={\displaystyle -\frac{\mathrm{i}\hbar}{2m}\left[\Psi^{*}\nabla\Psi-\Psi\nabla\Psi^{*}\right]={\displaystyle \frac{1}{m}{\rm Re}\left\{ \Psi^{*}(-\mathrm{i}\hbar\nabla)\Psi\right\} ,}}
\end{array}\label{eq:3.18-1}
\end{equation}
where $P_{{\rm tot}}=|\psi_{1}+\psi_{2}|^{2}=:|\Psi|^{2}$.

Eq.~(\ref{eq:vtot_fin-1}) has been derived for one particle in an
$n$-slit system. However, it is straightforward to extend this derivation
to the many-particle case. Due to the purely additive terms in the
expressions for the total current and total probability density, respectively,
also for \emph{N} particles, the only difference now is that the currents'
nabla operators have to be applied at all of the locations of the
respective \emph{N} particles, thus providing the quantum mechanical
formula
\begin{equation}
{\displaystyle \mathbf{J}_{{\rm tot}}}\left(N\right)={\displaystyle \sum_{i=1}^{N}}\frac{1}{m_{i}}{\rm Re}\left\{ \Psi\left(t\right)^{*}(-\mathrm{i}\hbar\nabla_{i})\Psi\left(t\right)\right\} ,
\end{equation}
where $\Psi\left(t\right)$ now is the total $N$-particle wave function,
whereas the total velocity fields are given by
\begin{equation}
\VEC v_{i}\left(t\right)=\frac{\hbar}{m_{i}}\mathrm{Im}\frac{\nabla_{i}\Psi\left(t\right)}{\Psi\left(t\right)}\;\forall i=1,...,N.
\end{equation}

Note that this result is similar in spirit to that of Norsen \emph{et~al.~}\cite{Norsen.2010theory,Norsen.2014can}
who with the introduction of a \emph{conditional wave function} $\tilde{\psi}_{i}$,
as opposed to the configuration-space wave function $\Psi$, rewrite
the guidance formula, for each particle, in terms of the $\tilde{\psi}_{i}$:

\begin{equation}
\frac{\d X_{i}\left(t\right)}{\d t}=\frac{\hbar}{m_{i}}\mathrm{Im}\left.\frac{\nabla\Psi}{\Psi}\right|_{\mathbf{\boldsymbol{x}=\boldsymbol{X}\left(t\right)}}\equiv\frac{\hbar}{m_{i}}\mathrm{Im}\left.\frac{\nabla\tilde{\psi}_{i}}{\tilde{\psi}_{i}}\right|_{x=X_{i}\left(t\right)},
\end{equation}
where the $X_{i}$ denote the location of one specific particle and
$\mathbf{X}\left(t\right)=\left\{ X_{1}\left(t\right),...,X_{N}\left(t\right)\right\} $
the actual configuration point. Thus, in this approach, each $\tilde{\psi}_{i}$
can be regarded as a wave propagating in physical 3-dimensional space.

In sum, with our introduction of a conditional probability $P(\VEC w_{i})$
for channels $\VEC w_{i}$, which include subquantum velocity fields,
we obtain the guidance formula also for $N$-particle systems. \textsl{Therefore,
what looks like the necessity in the dBB theory to superpose wave
functions in configuration space in order to provide an ``indigestible''
guiding wave, can equally be obtained by superpositions of all relational
amplitude configurations of waves in real 3-dimensional space. }\textsl{\emph{The
central ingredient for this to be possible is to consider the }}\textsl{emergence}\textsl{\emph{
of the velocity field from the interplay of the totality of all of
the system's velocity channels. We have termed the framework of our
approach a ``superclassical'' one, because in it are combined classical
levels at vastly different scales, i.e.\ at the subquantum and the
macroscopic levels, respectively.\clearpage{}}}

\section{Identity of the emergent force on a particle modeled by a bouncer
system and the quantum force of the deBroglie--Bohm theory\label{sec:force}}

With the results of the foregoing Chapter, we can now return to and
resolve the problem discussed in Chapter~1 of the apparent incompatibility
between the Bohmian force upon a quantum particle and the force exerted
on a bouncing droplet as formulated by Richardson \emph{et~al}.~\cite{Richardson.2014analogy}.
In fact, already a first look at the bouncer/walker model of our group
provides a clear difference as compared to the hydrodynamical force
studied by Richardson \emph{et~al}. For, whereas the latter investigate
the effects of essentially a single bounce on the fluid surface and
the acceleration of the bouncer as a consequence of this interaction,
our bouncer/walker model for quantum particles involves a much more
complex dynamical scenario: We consider the effects of a huge number
of bounces, i.e.\ typically of the order of $\nicefrac{1}{\omega}$,
like approximately $10^{21}$ bounces per second of an electron, which
constitute effectively a ``heating up'' of the bouncer's surrounding,
i.e.\ the subquantum medium related to the zero-point energy field. 

Note that as soon as a microdynamics is assumed, the development of
heat fluxes is a logical necessity if the microdynamics is constrained
by some macroscopic boundaries like that of a slit system, for example.
As we have shown in some detail~\cite{Groessing.2010emergence},
the thermal field created by such a huge number of bounces in a slit
system leads to an emergent average behavior of particle trajectories
which is identified as anomalous, and more specifically as ballistic,
diffusion. As such, the particle trajectories exiting from, say, a
Gaussian slit behave exactly as if they were subject to a Bohmian
quantum force. We were also able to show that this applies also to
$n$-slit systems, such that one arrives at a subquantum modeling
of the emergent interference effects at $n$ slits whose predicted
average behavior is identical to that provided by the dBB theory.

It is then easily shown that the average force acting on a particle
in our model is the same as the Bohmian quantum force. For, due to
the identity of our emerging velocity field with the guidance formula,
and because they essentially differ only via the notations due to
different forms of bookkeeping, their respective time derivatives
must also be identical. Thus, from Eq.~(\ref{eq:vtot_fin-1}) one
obtains the particle acceleration field (using a one-particle scenario
for simplicity) in an $n$-slit system as
\begin{align}
a_{\mathrm{tot}}\left(t\right) & =\frac{\d\mathbf{v}_{{\rm tot}}}{\d t}=\frac{\d}{\d t}\left(\frac{{\displaystyle \sum_{i=1}^{3n}}\VEC w_{i}P(\VEC w_{i})}{{\displaystyle \sum_{i=1}^{3n}}P(\VEC w_{i})}\right)\nonumber \\
 & =\frac{1}{\left({\displaystyle \sum_{i=1}^{3n}}P(\VEC w_{i})\right)^{2}}\left\{ \vphantom{\frac{{\displaystyle \sum^{3n}}}{{\displaystyle \sum^{3n}}}}\sum_{i=1}^{3n}\left[P(\VEC w_{i})\frac{\d\VEC w_{i}}{\d t}+\VEC w_{i}\frac{\d P(\VEC w_{i})}{\d t}\right]\left({\displaystyle \sum_{i=1}^{3n}}P(\VEC w_{i})\right)\right.\label{eq:3.1}\\
 & \qquad\qquad\qquad\qquad\qquad\qquad\left.-\left({\displaystyle \sum_{i=1}^{3n}}\VEC w_{i}P(\VEC w_{i})\right)\left({\displaystyle \sum_{i=1}^{3n}\frac{\d P(\VEC w_{i})}{\d t}}\right)\vphantom{\frac{{\displaystyle \sum^{3n}}}{{\displaystyle \sum^{3n}}}}\right\} .\nonumber 
\end{align}
Note in particular that~(\ref{eq:3.1}) typically becomes infinite
for regions $\left(\mathbf{x},t\right)$ where $P_{\mathrm{tot}}={\displaystyle \sum_{i=1}^{3n}}P(\VEC w_{i})\rightarrow0$,
in accordance with the Bohmian picture.

From~(\ref{eq:3.1}) we see that even the acceleration of one particle
in an $n$-slit system is a highly complex affair, as it nonlocally
depends on all other accelerations and temporal changes in the probability
densities across the whole experimental setup! In other words, this
force is truly emergent, resulting from a huge amount of bouncer-medium
interactions, both locally and nonlocally. This is of course radically
different from the scenario studied by Richardson \emph{et~al}.\ where
the effect of only a single local bounce is compared with the quantum
force. From our new perspective, it is then hardly a surprise that
the comparison of the two respective forces provides distinctive differences.
However, as we just showed, with the emergent scenario proposed in
our model, complete agreement with the Bohmian quantum force is established.

\section{Choose your Poison: How to introduce Nonlocality in a Hydrodynamic-like
model for quantum systems?\label{sec:poison}}

As already mentioned in the introduction of this paper, purely classical
hydrodynamical models are manifestly local and thus inadequate tools
to explain quantum mechanical nonlocality. Although nonlocal correlations
may also be obtainable within hydrodynamical modeling~\cite{Brady.2013violation},
there is no way to also account for dynamical nonlocality~\cite{Tollaksen.2010quantum}
in this manner. So, as correctly observed by Richardson \emph{et~al.}~\cite{Richardson.2014analogy},
droplet-surface wave interaction scenarios are not enough to serve
as a full-fledged analogy of the distinctly nonlocal dBB theory, for
example.

The question thus arises how in our much more complex, but still ``hydrodynamic-like''
bouncer/walker model nonlocal, or nonlocal-like, effects can come
about. To answer this, one needs to consider in more detail how the
elements of our model are constructed, which finally provide an elegant
formula, Eq.~(\ref{eq:vtot_fin-1}), identical with the guidance
formula in a (for simplicity: one-particle) system with $n$ slits.
(As shown above, the extension to $N$ particles is straightforward.)
As we consider, without restriction of generality, the typical example
of Gaussian slits, we introduce the Gaussians in the usual way, with
$\sigma$ related to the slit width, for the probability density distributions
(which in our model coincide with ``heat'' distributions due to
the bouncers' stirring up of the vacuum) just behind the slit. The
important feature of these Gaussians is that we do not implement any
cutoff for the distributions, but maintain the long tails which actually
then extend across the whole experimental setup, even if these are
only very small and practically often negligible amplitudes in the
regions far away from the slit proper. As the emerging probability
density current is given by the denominator of Eq.~(\ref{eq:vtot-1}),
we see that in fact the product $R_{1}R_{2}$ may be negligibly small
for regions where only a long tail of one Gaussian overlaps with another
Gaussian, nevertheless the last term in~(\ref{eq:vtot-1}) can be
very large despite the smallness of $R_{1}$ or $R_{2}$. It is this
latter part which is responsible for the genuinely quantum-like nature
of the average momentum, i.e.\ for its nonlocal nature. This is similar
in the Bohmian picture, but here given a more direct physical meaning
in that this last term refers to a difference in diffusive currents
as explicitly formulated in the last term of~(\ref{eq:vtot-1}).
Because of the mixing of diffusion currents from both channels, we
call this decisive term in $\mathbf{J_{\mathrm{\mathit{\mathrm{tot}}}}=P_{{\rm tot}}\mathbf{v}_{{\rm tot}}}$
the ``entangling current''.\ \cite{Mesa.2013variable}

Thus, one sees that formally one obtains genuine quantum mechanical
nonlocality in a hydrodynamic-like model with one particular ``unusual''
characteristic: the extremely feeble but long tails of (Gaussian or
other) distribution functions for probability densities exiting from
a slit extend nonlocally across the whole experimental setup. So,
we have nonlocality by explicitly putting it into our model. After
all, if the world is nonlocal, it would not make much sense to attempt
its reconstruction with purely local means. Still, so far we have
just stated a formal reason for how nonlocality may come about. Somewhere
in any theory, so it seems, one has to ``choose one's poison'' that
would provide nonlocality in the end. But what would be a truly ``digestible''
physical explanation? Here is where at present only some speculative
clues can be given.

For one thing, strict nonlocality in the sense of absolute simultaneity
of space-like separated events can never be proven in any realistic
experiment, because infinite precision is not attainable. This means,
however, that very short time lapses must be admitted in any operational
scenario, with two basic options remaining: i) either there is a time
lapse due to the finitely fast ``switching'' of experimental arrangements
in combination with instantaneous information transfer (but not signaling;
see Walleczek and Grössing, {[}forthcoming{]}), or ii) the information
transfer itself is not instantaneous, but happens at very high speeds
$v\ggg c$. 

How, then, can the implementation of nonlocal or nonlocal-like processes
with speeds $v\ggg c$ be argued for in the context of a hydrodynamic-like
bouncer/walker model? We briefly mention two options here. Firstly,
one can imagine that the ``medium'' we employ in our model is characterized
by oscillations of the zero-point energy throughout space, i.e.\ between
any two or more macroscopic boundaries as given by experimental setups.
Between these boundaries standing wave configurations may emerge (similar
to the Paris group's experiments, but now explicitly extending synchronously
over nonlocal distances). Here it might be helpful to remind ourselves
that we deal with solutions of the diffusion (heat conduction) equation.
At least (but perhaps only) formally, any change of the boundary conditions
is effective ``instantaneously'' across the whole setup. Alternatively,
if the experimental setup is changed such that old boundary conditions
are substituted by new ones, due to the all-space-pervading zero-point
energy oscillations, one ``immediately'' (i.e.\ after a very short
time of the order $t\sim\frac{1}{\omega}$) obtains a new standing
wave configuration that now effectively implies an almost instantaneous
change of probability density distributions, or relative phase changes,
for example. The latter would then become ``immediately'' effective
in that changed phase information is available across the whole domain
of the extended probability density distribution. We have referred
to this state of affairs as ``systemic nonlocality''~\cite{Groessing.2013dice}.
So, one may speculate that it is something like ``eigenvalues''
of the universe's network of zero-point fluctuations that may be responsible
for quantum mechanical nonlocality-eigenvalues which (almost?) instantaneously
change whenever the boundary conditions are changed.

A second option even more explicitly refers to the universe as a whole,
or, more particularly, to spacetime itself. If spacetime is an emergent
phenomenon as some recent work suggests~\cite{Padmanabhan.2014general},
this would very likely have strong implications for the modeling and
understanding of quantum phenomena. Just as in our model of an emergent
quantum mechanics we consider quantum theory as a possible limiting
case of a deeper level theory, present-day relativity and concepts
of spacetime may be approximations of, and emergent from a superclassical,
deeper level theory of gravity and/or spacetime. It is thus a potentially
fruitful task to bring both attempts together in the near future.
\begin{acknowledgments}
We thank Jan Walleczek for many enlightening discussions, and the
Fetzer Franklin Fund for partial support of the current work.
\end{acknowledgments}

\providecommand{\href}[2]{#2}\begingroup\raggedright\endgroup

\end{document}